\documentclass{article}
\usepackage{fleqn,ams}
\usepackage{amssymb}
\usepackage{epsfig}
\usepackage{verbatim}

\newcommand{\EFigure}[2]{%
\begin{center}\epsfig{file=#1,width=7cm,height=7cm}\\[12pt]
\refstepcounter{figure}Figure \thefigure: {\sl #2}\end{center}}
\begin{document}

\begin{center}
{\bf\Large  Diffusion model of evolution of superthermal
high-energy
particles under scaling in the early Universe}\\[12pt]
Yu.G.Ignatyev, R.A.Ziatdinov\\[12pt]
Kazan State Pedagogical University,\\ Mezhlauk str., 1, Kazan
420021, Russia
\end{center}

\begin{abstract}
The evolution of a superthermal relic component of matter is
studied on the basis of non-equilibrium model of Universe and the
Fokker-Planck type kinetic equation offered by one of the authors.
\end{abstract}

\section{Introduction}
The diffusion equation describing cosmological evolution of
superthermal particles under the assumption of
in\-te\-rac\-ti\-ons scaling recovery in the range of superhigh
energies was studied in paper \cite{Diffuze1} :
\begin{equation}\label{Yu53}
\frac{\partial {\cal G}}{\partial
\tau}=\frac{1}{x^2}\frac{\partial }{\partial
x}x^2\left(\frac{\partial {\cal G}}{\partial x}+ 2b(\tau){\cal G}
\right),
\end{equation}
where the dimensionless function $b(\tau)$ is determined as:
\begin{equation}\label{beta}
b(\tau)=\int\limits_0^\infty {\cal G}(\tau,x)x dx;
\end{equation}
It is necessary to solve the equation (\ref{Yu53}) with initial
and boundary conditions of the form:
\begin{equation}\label{Yu54}
{\cal G}(0,x)=G(x);
\end{equation}
\begin{equation}\label{Yu540}
\lim\limits_{x\to \infty}G(\tau,x)x^3=0,
\end{equation}
moreover the function $G(x)$ must satisfy the integral conditions:
\begin{equation} \label{Yu54a}
\int\limits_0^\infty G(x)x^2dx=1;
\end{equation}
\begin{equation} \label{Yu54b}
\int\limits_0^\infty G(x)x^3dx=1.
\end{equation}
As the detailed researches have shown, the solution of kinetic
equation (\ref{Yu53}) in case of $b(\tau)=0$ obtained in
\cite{Diffuze1} becomes correct only at greater times of
evolution, therefore it becomes problematical to establish
parameters relationship of this solution with the initial
distribution. In this paper we discuss the evolution of system at
small times.

It is necessary to note that the exact solution of the kinetic
equation in diffusive approximation (\ref{Yu53}), satisfying the
normalization relations (\ref{Yu54a}), (\ref{Yu54b}), is an
equilibrium ulatrarelativistic Bolzman distribution with conformal
temperature $\tau=\frac {1}{3}$:
\begin{equation} \label{Yu54c}
f_0=\frac {27}{2} e^{-3x}.
\end{equation}

\section{Numerical model of the initial distribution}

Let's consider the initial distribution analogous to Fermi-Dirac
distribution, given in the form of infinitely
dif\-fe\-ren\-ti\-ab\-le and integrable function:
\begin{equation} \label{Yu55}
G_0(x)=\frac {A}{e^{\xi x-y}+1},
\end{equation}
where $A$,$\xi$,$y$ - parameters of the initial distribution.
These parameters must be such that normalization relations
(\ref{Yu54a}),(\ref{Yu54b}) fulfill automatically. Thus, we have
two algebraic relations on three parameters, solving which through
parameter $y$ we find:
\begin{equation} \label{Yu56}
\xi(y)=\frac {{\displaystyle \int\limits_0^\infty \frac
{t^3dt}{e^{(t-y)}+1}}}{{\displaystyle\int\limits_0^\infty \frac
{t^2dt}{e^{(t-y)}+1}}}; \quad A(y)=\frac
{\xi^3(y)}{{\displaystyle\int\limits_0^\infty \frac
{t^2dt}{e^{(t-y)}+1}}}
\end{equation}
Slow convergence of improper integrals in (\ref{Yu56}) leads to
necessity of their transformation to following form that is more
suitable for numeric calculations:
$$J_1(y)=\int\limits_0^\infty \frac{t dt}{e^{(t-y)}+1}\equiv
$$
\begin{equation}\label{In0}
y\int\limits_0^\infty\frac{dx}{e^x+1}+\int\limits_0^\infty\frac{xdx}{e^x+1}+y\int\limits_0^y\frac{dx}{e^{-x}+1}-
\int\limits_0^y\frac{xdx}{e^x+1}.
\end{equation}
$$J_2(y)=\int\limits_0^\infty \frac{t^2 dt}{e^{(t-y)}+1}\equiv
y^2\int\limits_0^\infty\frac{dx}{e^x+1}+$$
\begin{equation}\label{In1}
2y\int\limits_0^\infty\frac{xdx}{e^x+1}+\int\limits_0^\infty\frac{x^2dx}{e^x+1}+y^2\int\limits_0^y\frac{dx}{e^{-x}+1}-
2y\int\limits_0^y\frac{xdx}{e^x+1}+
\int\limits_0^y\frac{x^2dx}{e^x+1}.
\end{equation}
$$J_3(y)=\int\limits_0^\infty \frac{t^3 dt}{e^{(t-y)}+1}\equiv
y^3\int\limits_0^\infty\frac{dx}{e^x+1}+3y^2\int\limits_0^\infty\frac{xdx}{e^x+1}
$$
$$
+3y\int\limits_0^\infty\frac{x^2dx}{e^x+1}
+\int\limits_0^\infty\frac{x^3dx}{e^x+1}+y^3\int\limits_0^y\frac{dx}{e^{-x}+1}-
$$
\begin{equation}\label{In2}
3y^2\int\limits_0^y\frac{xdx}{e^x+1}+
3y\int\limits_0^y\frac{x^2dx}{e^x+1}-\int\limits_0^y\frac{x^3dx}{e^x+1}.
\end{equation}
Using known representations of Riemann $\zeta$ - functions and
Bernoulli numbers (e.g., see \cite{Grad}):
\begin{equation}\label{In2a}
\int\limits_0^\infty
\frac{x^{n-1}}{e^x+1}dx=(1-2^{1-n})\Gamma(n)\zeta(n);
\end{equation}
\begin{equation}\label{In3}
\int\limits_0^\infty
\frac{x^{2n-1}}{e^x+1}dx=\frac{2^{2n-1}-1}{2n}\pi^{2n}B_n
\end{equation}
and introducing dimensionless functions:
\begin{equation}\label{In4}
S(n,y)=\int\limits_0^y\frac{x^n}{e^{-x}+1}dx,\quad (n=0,1,2,...),
\end{equation}
($S(0,y)=\ln(1+e^y)/2$) we get expressions for
(\ref{In0})-(\ref{In2}):
\begin{equation}\label{In5_0}
\quad J_1(y)=y\ln 2+\frac{\pi^2}{12}+yS(0,y)-S(1,y);
\end{equation}
$$
J_2(y)=y^2\ln 2+y\frac{\pi^2}{6}+\frac{3}{2}\zeta(3)+y^2S(0,y)-
$$
\begin{equation}\label{In5}
2yS(1,y)+S(2,y);
\end{equation}
$$
J_3(y)=y^3\ln
2+y^2\frac{\pi^2}{4}+y\frac{9}{2}\zeta(3)+\frac{7}{120}\pi^4+
$$
\begin{equation}\label{In6}
y^3S(0,y)-3y^2S(1,y)+3yS(2,y)-S(3,y).
\end{equation}
Thus, by means of introduced functions (\ref{In0})-(\ref{In2}) we
find expression for $b(\tau)$:
$$
b_0(y)=A(y)\int\limits_0^\infty \frac{xdx}{e^{\xi(y)x-y}+1}\equiv
$$
\begin{equation}\label{In7}
\frac{A(y)}{\xi^2(y)}\int\limits_0^\infty
\frac{xdx}{e^{x-y}+1}=\frac{J_3(y)J_1(y)}{J^2_2(y)}.
\end{equation}
Presentation of (\ref{Yu56}) by means of $S(n,y)$ and Riemann
$\zeta$ - functions makes numeric computations simpler. The
results of integration are shown on Figure 1,2 and we see that
$\xi(y)$ is monotone increasing, $A(y)$ is monotone decreasing
function.

\EFigure{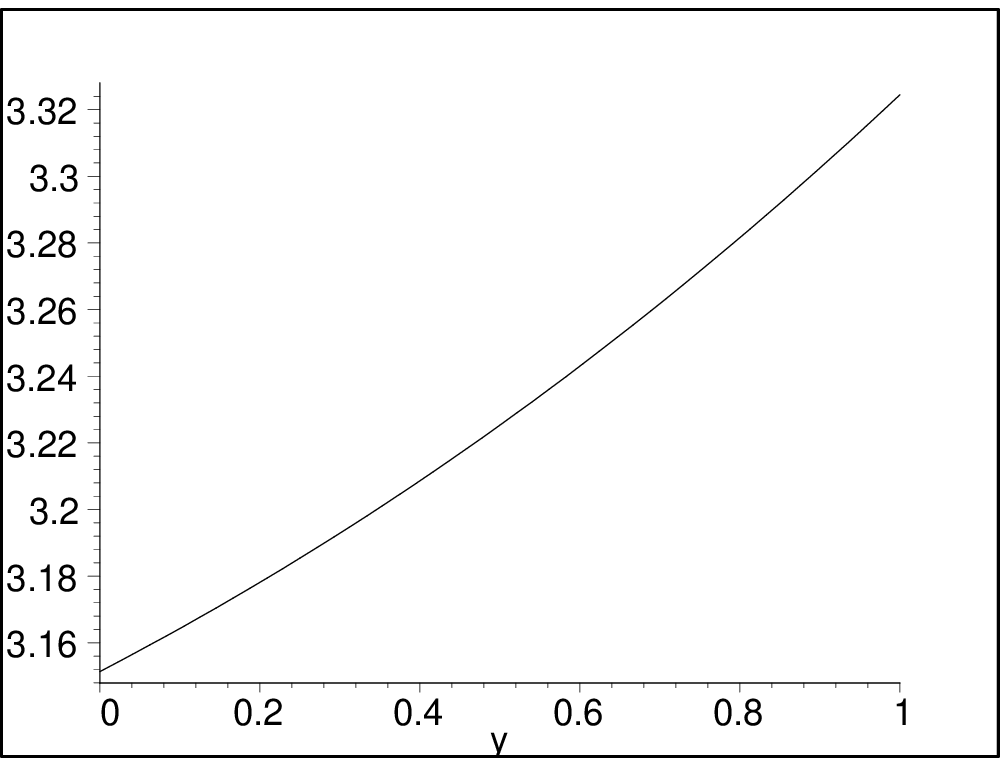}{\label{pic01.eps}Plot of $\xi(y)$ function.}

\EFigure{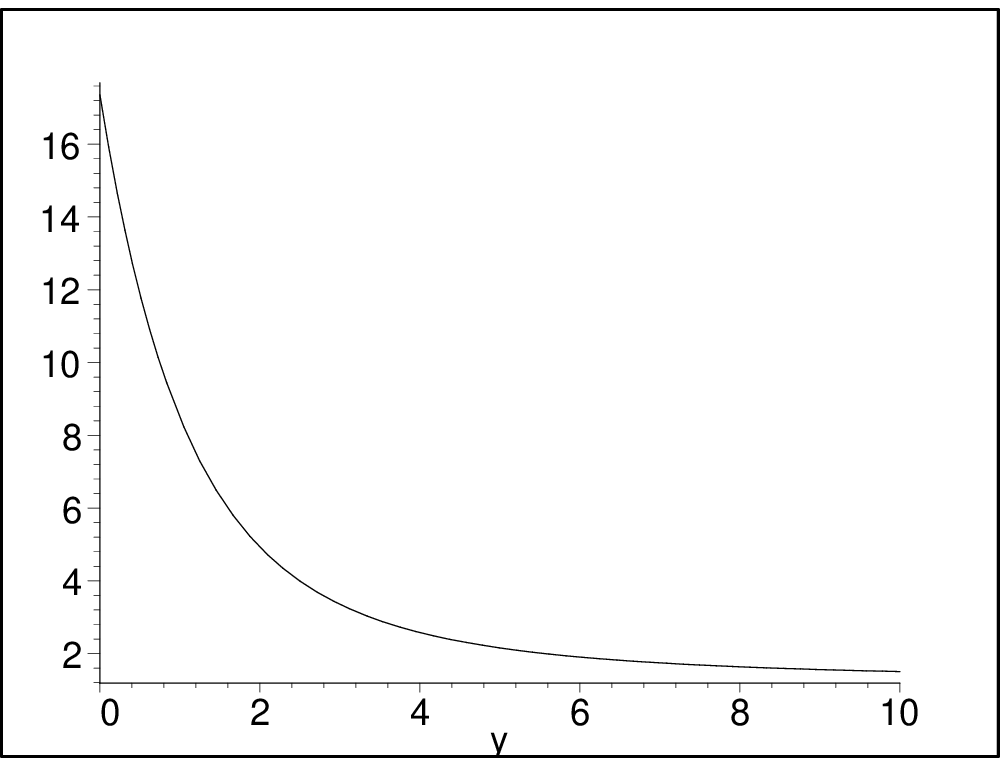}{\label{pic02.eps}Plot of $A(y)$ function.}
As a result, normalized initial distribution function is defined
by one arbitrary parameter, $y$:
\begin{equation} \label{Yu57}
G_0(x,y)=\frac {A(y)}{e^{\xi(y) x-y}+1},
\end{equation}
which controls the the degree of non-equilibrium of initial
distribution (\ref{Yu55}), Figure 3.

\EFigure{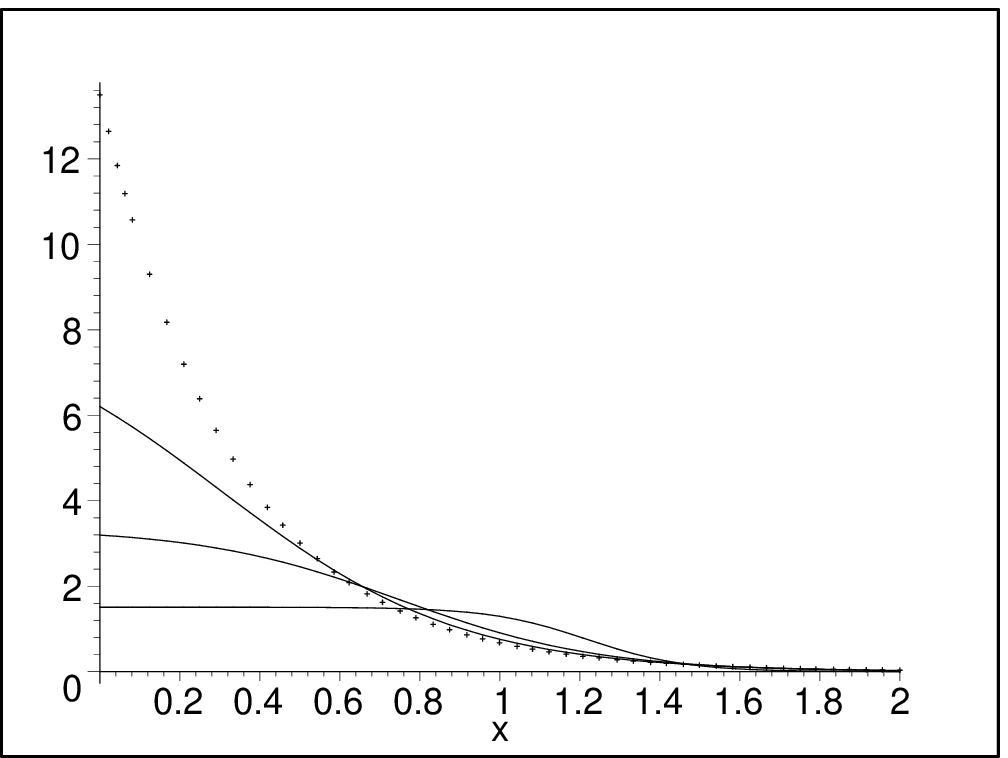}{\label{pic03.eps}Normalized initial
distribution $G_0(x,y)$ according to parameter $y$: top-down:
$y=1$, $y=3$, $y=10$. The initial distribution (\ref{Yu55}) is
shown by dotted line.}

\section{Diffusion equation and integral conservation laws}
As known (see, e.g. \cite{Yu3}), the strict consequences from
general relativistic kinetic theory in case of elastic
col\-li\-si\-ons are integral conservation laws of particles total
number and their energy. The solution of diffusion equation
(\ref{Yu53}), obtained on the basis of general relativistic
kinetic equations, also must satisfy these laws. Therefore it is
necessary to verify this fact.

Multiplying by $x^2$ the both sides of equation (\ref{Yu53}) and
integrating received expression in parts on all interval of
impulsive variable $x$, we get:
\begin{equation} \label{YuZi1}
\frac {d}{d\tau}\int\limits_0^\infty x^2 G(x,\tau)dx=
\left.\left[x^2\left(\frac{\partial G}{\partial
x}+2b(\tau)G\right)\right]\right|_0^\infty.
\end{equation}
Assuming later on:
\begin{equation}\label{YuZi2}
\quad \lim\limits_{x\to 0}x^2 G; \quad \lim\limits_{x\to
0}x^2\frac{\partial G}{\partial x}=0;
\end{equation}
\begin{equation}\label{YuZi3}
\quad \lim\limits_{x\to \infty}x^2 G=0; \quad \lim\limits_{x\to
\infty}x^2 \frac{\partial G}{\partial x}=0;
\end{equation}
we obtain from (\ref{YuZi1}):
\begin{equation}\label{YuZi3}
\int\limits_0^\infty x^2 G(x,\tau)dx=\rm{Const}.
\end{equation}
According to (\ref{Yu54a}) the constant at the right side of
(\ref{YuZi3}) equals to 1.

Multiplying now by $x^3$ the both sides of (\ref{Yu53}) and
integrating received expression in parts on all interval, we get:
$$
\frac {d}{d\tau}\int\limits_0^\infty x^3 G(x,\tau)dx=
\left.\left[x^3\left(\frac{\partial G}{\partial
x}+2b(\tau)G\right)\right]\right|_0^\infty-
$$
\begin{equation} \label{YuZi4}
-\int\limits_0^\infty x^2 \left(\frac{\partial G}{\partial
x}+2b(\tau)G\right)dx
\end{equation}
Considering (\ref{YuZi2}), (\ref{YuZi3}), (\ref{Yu540}) and once
more integrating received expression in parts, we find:
$$
\frac {d}{d\tau}\int\limits_0^\infty x^3 G(x,\tau)dx=
-\left.x^2G\right|_0^\infty+
$$
\begin{equation} \label{YuZi7}
+2\int\limits_0^\infty xG dx-2b(\tau)\int\limits_0^\infty x^2Gdx.
\end{equation}
Taking into account relations (\ref{beta}),(\ref{Yu54a}), and also
(\ref{YuZi2}),(\ref{YuZi3}), we finally obtain from (\ref{YuZi7}):
\begin{equation}\label{YuZi8}
\int\limits_0^\infty x^3 G(x,\tau)dx=\rm{Const}.
\end{equation}
According to (\ref{Yu54b}) the constant at the right side of
(\ref{YuZi8}) equals to 1.

It is enough in order to $G(x)$ function's degree of magnitude
satisfy the following strong inequalities for realization of
relations (\ref{YuZi2}),(\ref{YuZi3}),(\ref{Yu540}):
\begin{equation}\label{YuZi8a}
G(x)|_{x\rightarrow 0}<\frac {1}{x},\quad
G(x)|_{x\rightarrow\infty}<\frac {1}{x^3}.
\end{equation}
We also note that realization of the energy conservation law is
provided with the presence of term with $b(\tau)$ coefficient at
the right side of diffusion equation (\ref{Yu53}). If this term is
missing, the process of energy transmitting at small times is not
considered.
\section{Expansion of diffusion equation by smallness of $\tau$}
This approximation match up the early stages of universe
evolution, when particle interactions are inessential:
\begin{equation}\label{YuZi9}
\tau \ll 1.
\end{equation}
Then the term at the left side of diffusion equation is main and
taking into account the initial distribution (\ref{Yu54}) we have:
\begin{equation}\label{YuZi10}
\frac{\partial G}{\partial \tau}=0, \Rightarrow G(x,\tau)=G_0(x).
\end{equation}
Expanding the right side of equation (\ref{Yu53}), substituting
expression (\ref{YuZi10}) as the initial distribution and
integrating in time variable we obtain the first correction.
Sequentially iterating this procedure we get the recurring formula
for finding higher approximations:\footnote{Here and further, if
not mentioned especially, $k\in \mathbb{N}$.}
$$G_{k+1} =\int \frac{1}{x^{2} }\frac{\partial } {\partial x} x^{2}
\left[\frac{\partial G_{k}}{\partial x}+\right.$$
\begin{equation}\label{YuZi22}
\left. 2\sum _{i=0}^{k }\left(G_{i} \int _{0}^{\infty}G_{k-i}xdx
\right) \right]d\tau,
\end{equation}
Considering that correction of $k$-order is proportional to
$\tau^k$ and integrating (\ref{YuZi22}), we get:
$$G_{k+1} =\frac{\tau^{k+1}}{k+1}\frac{1}{x^{2} }\frac{\partial } {\partial
x} x^{2} \left[\frac{\partial g_{k}}{\partial x}+\right.$$
\begin{equation}\label{YuZi24}
\left.2\sum _{i=0}^{k }\left(g_{i} \int _{0}^{\infty}g_{k-i}xdx
\right) \right].
\end{equation}
Differentiating in expression (\ref{YuZi24}), we finally get:
$$G_{k+1}=\frac{\tau^{k+1}}{k+1}\left\{\frac{\partial^2 g_k}{\partial
x^2}+\frac{2}{x}\frac{\partial g_k}{\partial x}+\right.$$
\begin{equation}\label{YuZi26}
\left.2\sum _{i=0}^{k }\left[\left(\frac{\partial g_i}{\partial
x}+\frac {2g_i}{x}\right)b_{k-i}\right]\right\} .
\end{equation}
We also find the recurring formula for determining function
$b(\tau)$ in equation $(\ref{Yu53})$. Considering $(\ref{YuZi24})$
we obtain according to $(\ref{beta})$:
$$b_{k+1}=\frac{\tau^{k+1}}{k+1}\int _{0}^{\infty}\frac{dx}{x}\frac{\partial
} {\partial x} x^{2} \left[\frac{\partial g_{k}}{\partial
x}+\right.$$
\begin{equation}\label{YuZi25a}
\left.2\sum _{i=0}^{k }\left(g_{i} \int _{0}^{\infty}g_{k-i}xdx
\right) \right].
\end{equation}
Integrating in parts an integral in (\ref{YuZi25a}) and supposing
henceforward:
\begin{equation}\label{YuZi25c}
\quad \lim\limits_{x\to 0}x g_i=0; \quad \lim\limits_{x\to 0}x
\frac{\partial g_k}{\partial x}=0;
\end{equation}
\begin{equation}\label{YuZi25d}
\quad \lim\limits_{x\to \infty}x g_i=0; \quad \lim\limits_{x\to
\infty}x \frac{\partial g_k}{\partial x}=0.
\end{equation}
we get finally:
$$b_{k+1}=$$
\begin{equation}\label{YuZi25e}
\frac{\tau^{k+1}}{k+1}\int _{0}^{\infty}\left[\frac{\partial
g_{k}}{\partial x} +2\sum _{i=0}^{k }\left(g_{i} b_{k-i} \right)
\right] dx.
\end{equation}
Now we prove that the corrections to normalized initial
distribution (\ref{Yu55}) can not change the total number density
and energy density at every step of iterations.

Multiplying by $x^2$ the both sides of (\ref{YuZi24}) and
in\-te\-gra\-ting on all interval, we get:
$$\int\limits_0^\infty x^2G_{k+1}dx=\frac{\tau^{k+1}}{k+1}\left
\{x^2\left[\frac{\partial g_{k}}{\partial x}+\right. \right. $$
\begin{equation}\label{YuZi27}
\left.\left. \left.2\sum _{i=0}^{k }\left(g_{i} \int
_{0}^{\infty}g_{k-i}xdx \right) \right]\right\}\right|_0^\infty.
\end{equation}
Supposing henceforward:
\begin{equation}\label{YuZi28}
\quad \lim\limits_{x\to 0}x^2 g_i=0; \quad \lim\limits_{x\to 0}x^2
\frac{\partial g_k}{\partial x}=0;
\end{equation}
\begin{equation}\label{YuZi29}
\quad \lim\limits_{x\to \infty}x^2 g_i=0; \quad \lim\limits_{x\to
\infty}x^2 \frac{\partial g_k}{\partial x}=0.
\end{equation}
we obtain from (\ref{YuZi27}):
\begin{equation}\label{YuZi30}
\int\limits_0^\infty x^2 G_{k+1}dx=0.
\end{equation}
Multiplying by $x^3$ the both sides of (\ref{YuZi24}) and
integrating on all interval, we get:
$$\int\limits_0^\infty x^3 G_{k+1}dx=$$
$$
=\frac{\tau^{k+1}}{k+1}\left.\left\{x^3\left[\frac{\partial
g_{k}}{\partial x} +2\sum _{i=0}^{k }\left(g_{i} \int
_{0}^{\infty}g_{k-i}x dx \right) \right] \right\}
\right|_0^\infty-
$$
\begin{equation} \label{YuZi31}
\frac{\tau^{k+1}}{k+1}\int\limits_0^\infty x^2
\left[\frac{\partial g_{k}}{\partial x} +2\sum _{i=0}^{k
}\left(g_{i} \int _{0}^{\infty}g_{k-i}x dx \right) \right]dx.
\end{equation}
Considering (\ref{YuZi28}), (\ref{YuZi29}), (\ref{YuZi8}) and once
more integrating received expression in parts, we find from
(\ref{YuZi31}):
$$
\int\limits_0^\infty x^3 G_{k+1}dx= \frac{\tau^{k+1}}{k+1}\left \{
-\left.x^2g_k\right|_0^\infty+2\int\limits_0^\infty x g_k
dx-\right.
$$
\begin{equation} \label{YuZi34}
\left. -2\int\limits_0^\infty x^2 \sum _{i=0}^{k }\left(g_{i} \int
_{0}^{\infty}g_{k-i}x dx \right) dx\right \}.
\end{equation}
After an obvious simplifications
$$
\int\limits_0^\infty x^3 G_{k+1}dx= \frac{\tau^{k+1}}{k+1}\left \{
-\left.x^2g_k\right|_0^\infty+2\int\limits_0^\infty x g_k
dx-\right.
$$
$$
-2\int\limits_0^\infty x g_k dx\int\limits_0^\infty x^2 g_{0} dx-
$$
\begin{equation} \label{YuZi35}
\left.-2\sum _{i=1}^{k }\left (\int\limits_0^\infty g_{k-i}x dx
\int _{0}^{\infty} x^2 g_{i}dx \right ) \right \},
\end{equation}
and taking into account expressions (\ref{Yu54a}), (\ref{YuZi28}),
(\ref{YuZi29}), (\ref{YuZi30}), we get from (\ref{YuZi35}) at
last:
\begin{equation}\label{YuZi36}
\int\limits_0^\infty x^3 G_{k+1}dx=0.
\end{equation}
Thus, we certain that the distribution function iterations of each
step don't change the total number of particles and energy, this
is useful tool for calculation correctness. It follows from
(\ref{YuZi36}) that the correction of any order is
alternating-sign on the interval $[0, +\infty)$. Hence, small time
$\tau$ approximation is completely equivalent to ex\-pan\-si\-on
of exact function $G(x,\tau)$ to Taylor series by $\tau$ powers.
\subsection{The first order approximation}
As a first approximation according to recurring formula
(\ref{YuZi24}) we have:
\begin{equation}\label{YuZi11a}
G_1=\tau g_1;
\end{equation}
\begin{equation}\label{YuZi11}
g_1=\frac{1}{x^2}\frac{\partial }{\partial
x}x^2\left(\frac{\partial G_0}{\partial x}+ 2b_0 G_0 \right).
\end{equation}
Substituting expression (\ref{YuZi11}) into (\ref{beta}) and
integrating in parts, also considering that function $G_0(x,y)$,
defined by expression (\ref{Yu55}), with its derivatives approach
to 0 at $x\to\infty$ faster than any of exponential functions and
at $x\rightarrow 0$ has finite derivatives, we obtain:
$$
b_1=\tau\int\limits_0^\infty\frac{dx}{x}\frac{\partial}{\partial
x}\left(\frac{\partial G_0}{\partial x}+2b_0G_0(x) \right)=$$
$$=\tau\left. \left[x\left(\frac{\partial G_0}{\partial x}+2b_0G_0(x)
\right)\right]\right|_0^\infty+$$

$$+\tau\int\limits_0^\infty\left(\frac{\partial G_0}{\partial x}+2b_0G_0(x) \right)
dx=$$

$$=-\tau G_0(0)+\tau 2b_0 \int\limits_0^\infty G_0dx\Rightarrow$$
$$b_1(\tau)= -\tau\frac{A(y)e^{y}}{e^{y}+1}+\tau 2b_0 \int\limits_0^\infty G_0dx.$$
Substituting here the initial distribution (\ref{Yu55}), we find:
\begin{equation}\label{YuZi13a}
\int\limits_0^\infty G_0dx=\frac{A(y)}{a(y)}\ln(1+e^y),
\end{equation}
therefore
\begin{equation}\label{YuZi13a}
b_1(\tau)=\tau
A(y)\left[\frac{2b_0\ln(1+e^y)}{a(y)}-\frac{e^{y}}{e^{y}+1}\right].
\end{equation}
Then, substituting the function $G_0(x,y)$ into (\ref{YuZi11}), we
determine an explicit form of linear approximation to the initial
distribution:
\EFigure{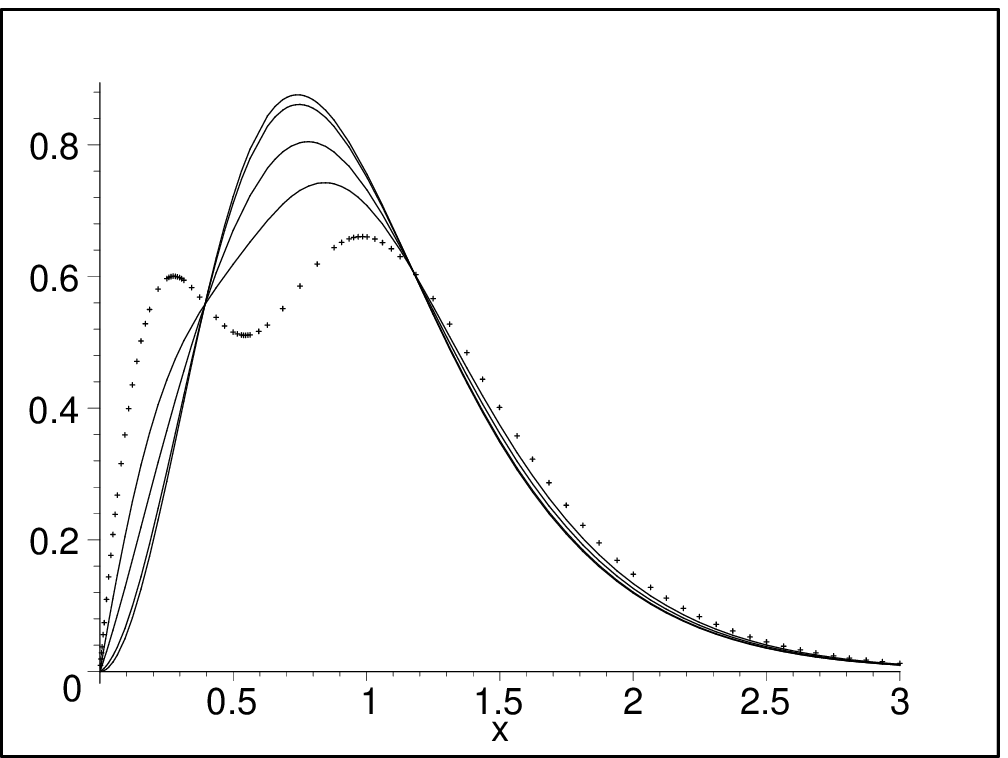}{\label{pic04.eps}Evolution of number density
of particles, $dn(\tau,x,y)=x^2G(\tau,x,y)$, as a first
approximation at $y=1$. Solid lines from left to right: $\tau=0;
0,01; 0,05; 0,1$, dotted line-$\tau=0,2$.}

\EFigure{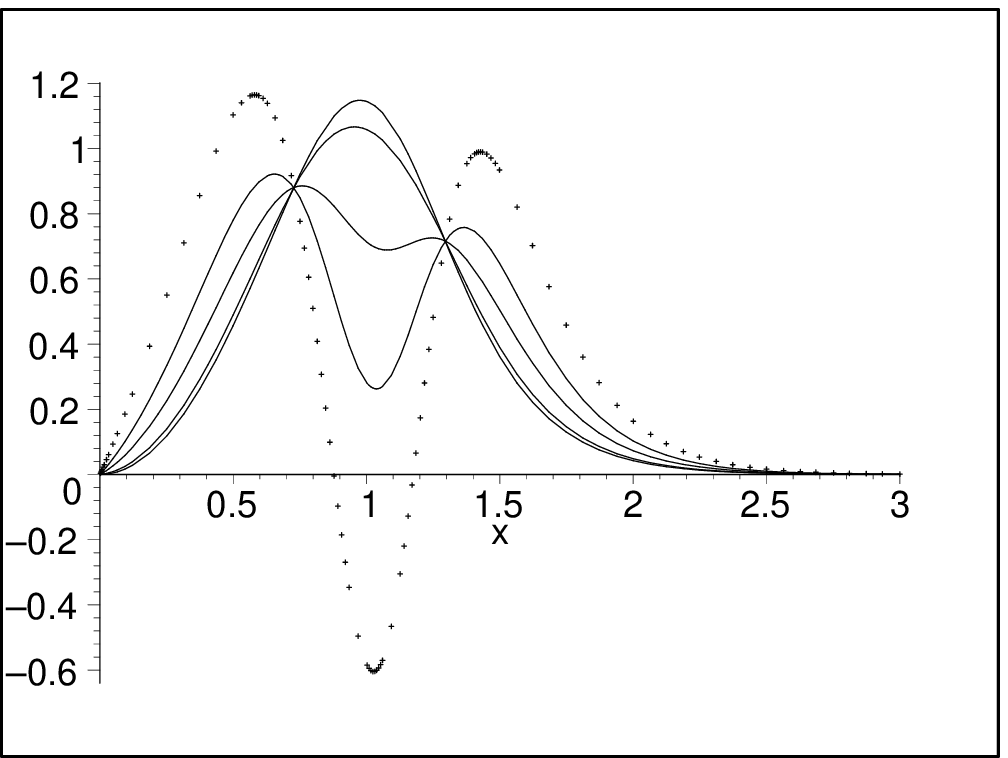}{\label{pic06.eps}Evolution of number density
of particles, $dn(\tau,x,y)=x^2G(\tau,x,y)$, as a first
approximation at $y=6$. Solid lines from left to right: $\tau=0;
0,01; 0,05; 0,1$, dotted line-$\tau=0,2$.}
$$g_{1} \left (x,y\right )=\frac {2 A(y)}{x} \frac
{e^{\xi(y)x-y}[2b_{0}(y)-\xi(y)]+2b_{0}(y)}{(e^{\xi(y)x-y}+1)^2}-$$
$$-\frac{A(y)\xi(y)[\xi(y)+2b_{0}(y)]e^{\xi(y)x-y}}{(e^{\xi(y)x-y}+1)^2}+$$

\begin{equation}\label{YuZi15}
+\frac {2A(y)\xi(y)^2e^{2(\xi(y)x-y)}}{(e^{\xi(y)x-y}+1)^3}.
\end{equation}
On the figures 4-11 is shown the evolution of distribution of
number density and their energy density as a first approximation
at the values of $y=1,3,6,10$, when the significant particles part
of the initial distribution lies in the area of small energy
values.

As we see from these and also following figures:

\begin{enumerate}
\item In number density and their energy density dis\-tri\-bu\-ti\-ons
always appear 2 maximums and 1 minimum, which shifts to area of
higher energy values with increasing of $y$ parameter;

\item Always appear such a moment of time, at which the minimum of
distribution function takes negative value;

\item After a time the first minimum shifts to the area of lower
energies, the second minimum - to the area of higher energies.

\end{enumerate}

\EFigure{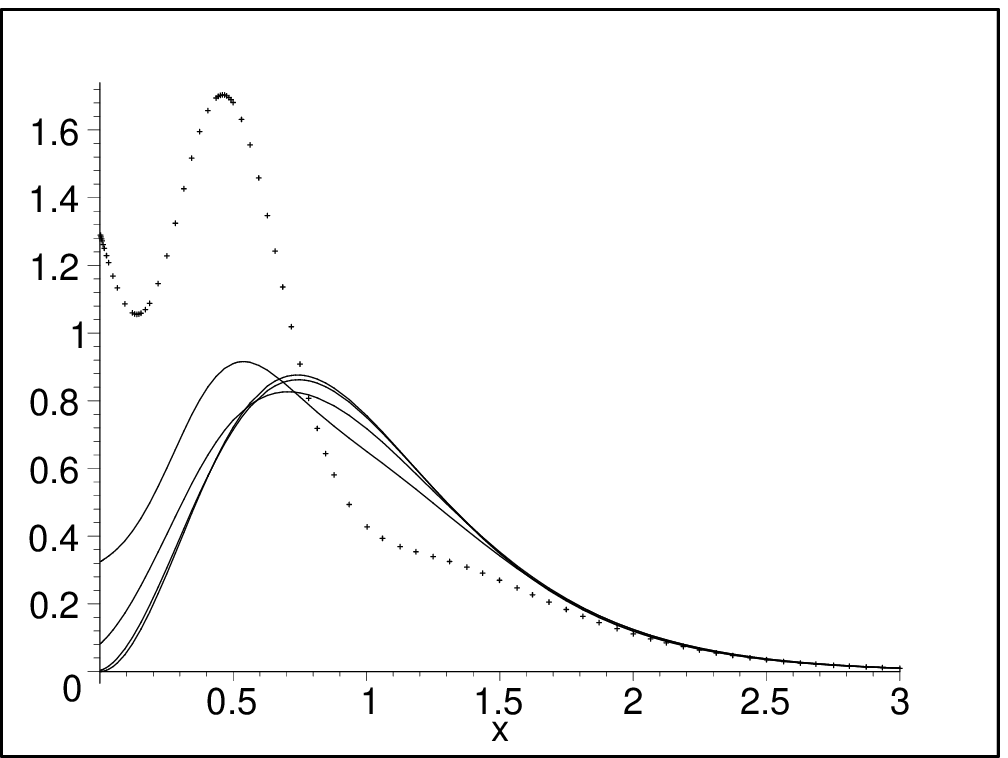}{\label{pic12.eps}Evolution of number density
of particles, $dn(\tau,x,y)=x^2G(\tau,x,y)$, as a second
approximation at $y=1$. Solid lines from left to right: $\tau=0;
0,01; 0,05; 0,1$, dotted line-$\tau=0,2$.}

\EFigure{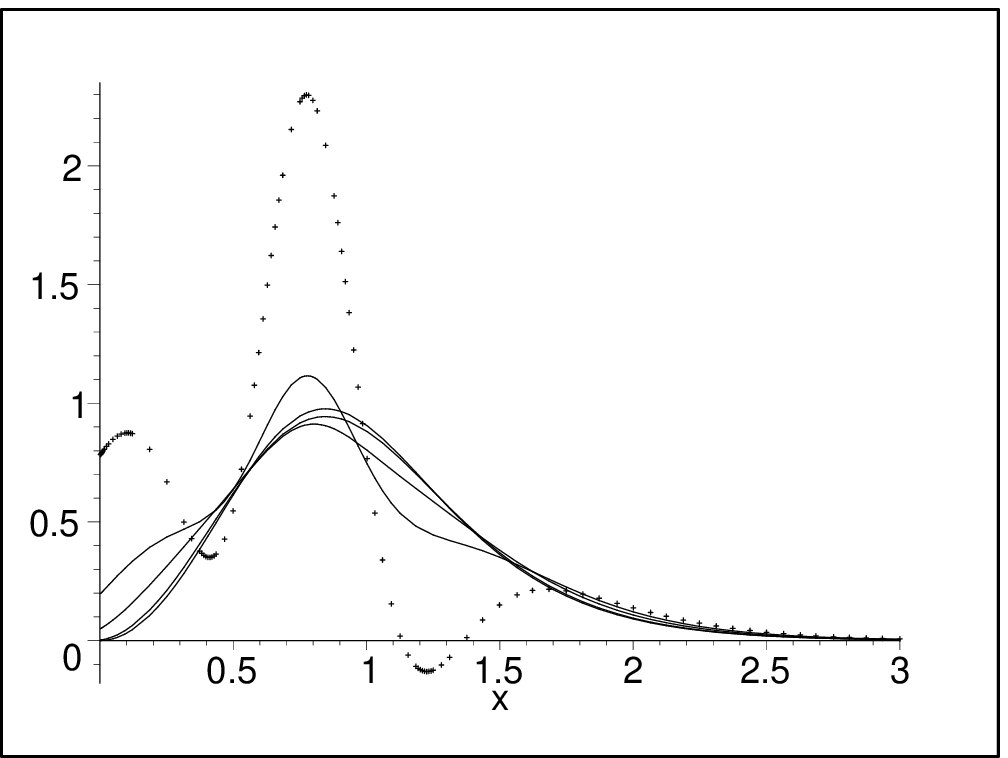}{\label{pic13.eps}Evolution of number density
of particles, $dn(\tau,x,y)=x^2G(\tau,x,y)$, as a second
approximation at $y=3$. Solid lines from left to right: $\tau=0;
0,01; 0,05; 0,1$, dotted line-$\tau=0,2$.}
\subsection{The second order approximation}

As a second approximation according to (\ref{YuZi24}) we have:
\begin{equation}\label{YuZi16}
G_{2} =\frac{\tau ^{2}}{2}g_{2};
\end{equation}
\begin{equation}\label{YuZi17}
g_{2}=\frac{1}{x^{2} } \frac{\partial } {\partial x} x^{2} \left
(\frac{\partial g_{1} }{\partial x} +2g_{1} b_{0} +2G_{0} b_{1}
\right).
\end{equation}
Substituting then $G_{0}$ and $g_{1}$ into (\ref{YuZi17}), we find
the explicit form of the second order approximation.\footnote {In
case that the explicit form of the second order approximation is
an unwieldy we don't cite it here}

\EFigure{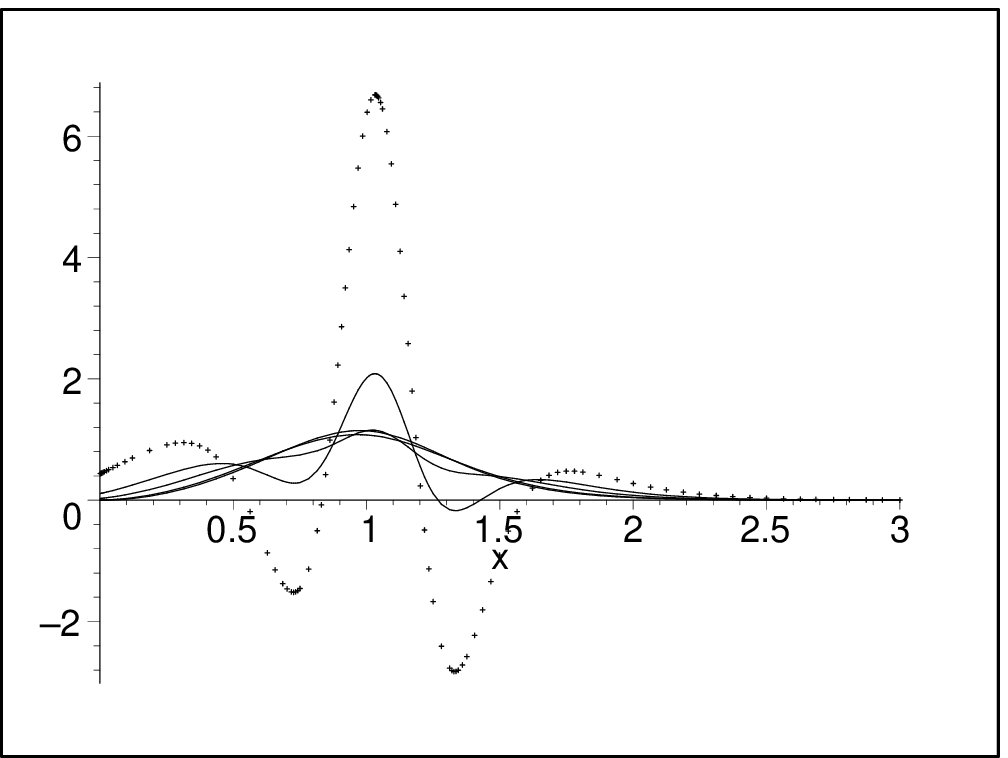}{\label{pic14.eps}Evolution of number density
of particles, $dn(\tau,x,y)=x^2G(\tau,x,y)$, as a second
approximation at $y=6$. Solid lines from left to right: $\tau=0;
0,01; 0,05; 0,1$, dotted line-$\tau=0,2$.}

\EFigure{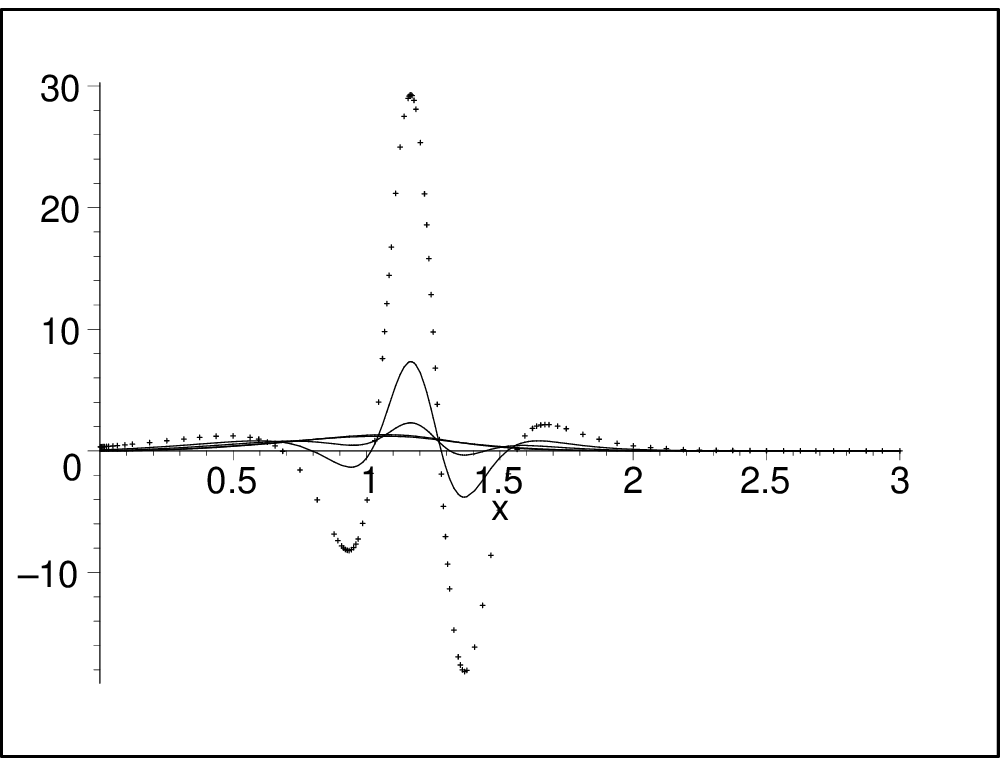}{\label{pic15.eps}Evolution of number density
of particles, $dn(\tau,x,y)=x^2G(\tau,x,y)$, as a second
approximation at $y=10$. Solid lines from left to right: $\tau=0;
0,01; 0,05; 0,1$, dotted line-$\tau=0,2$.}

\EFigure{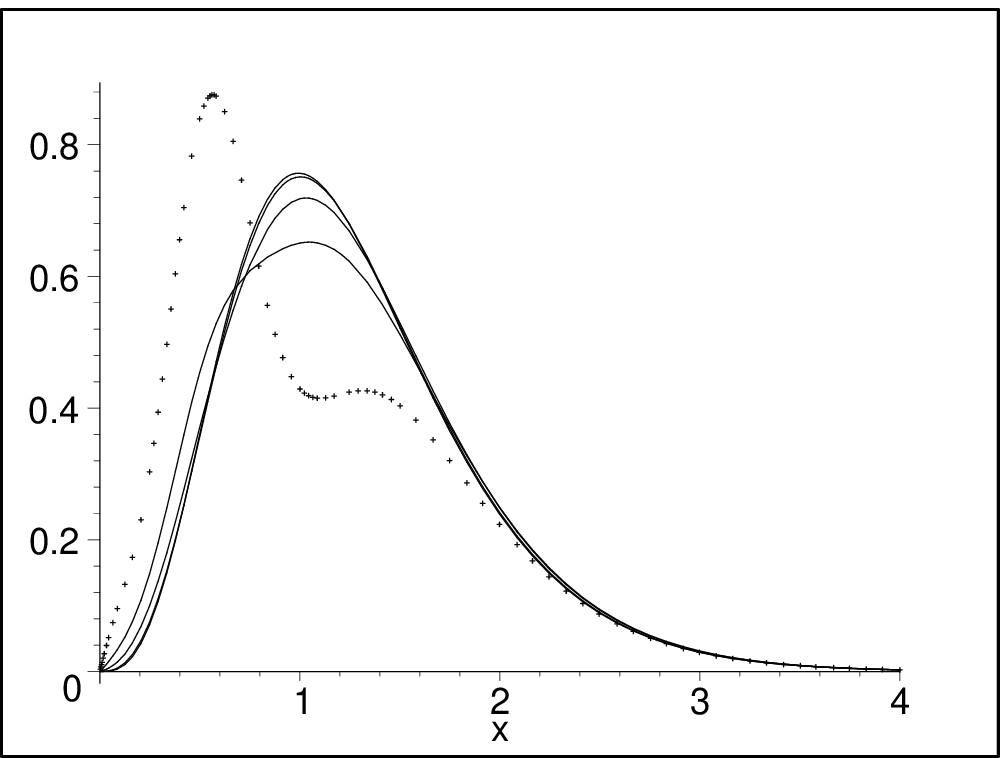}{\label{pic16.eps}Evolution of the energy
distribution of particles,
$d\varepsilon(\tau,x,y)=x^3G(\tau,x,y)$, as a second approximation
at $y=1$. Solid lines from left to right: $\tau=0; 0,01; 0,05;
0,1$, dotted line-$\tau=0,2$.}

\EFigure{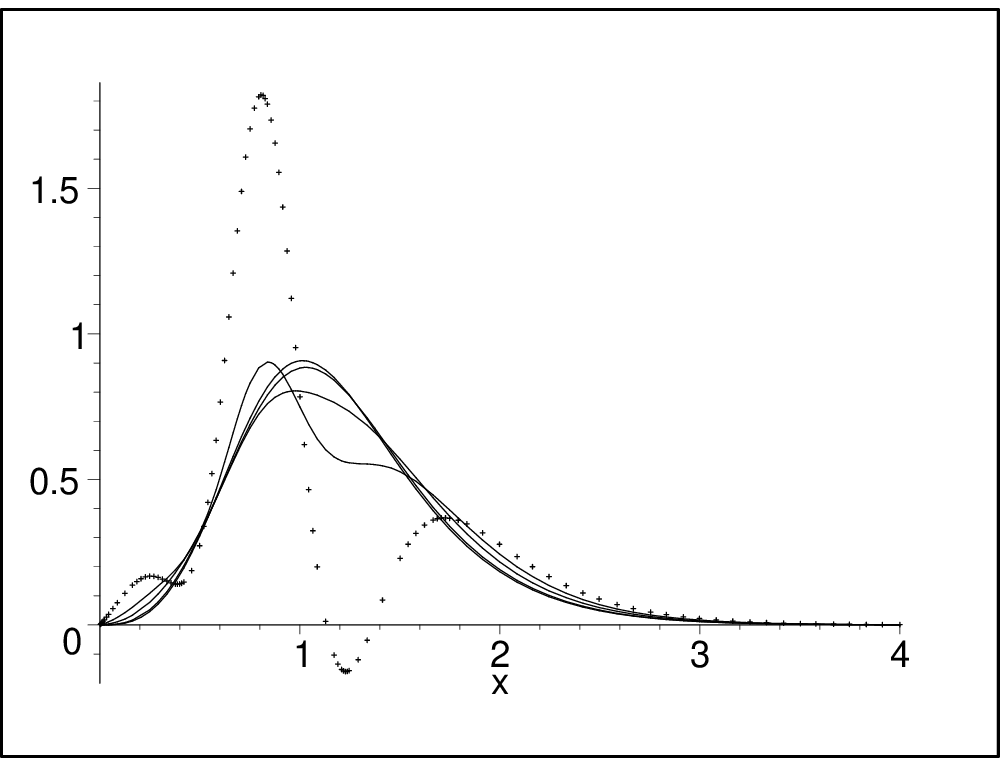}{\label{pic17.eps}Evolution of the energy
distribution of particles,
$d\varepsilon(\tau,x,y)=x^3G(\tau,x,y)$, as a second approximation
at $y=3$. Solid lines from left to right: $\tau=0; 0,01; 0,05;
0,1$, dotted line-$\tau=0,2$.}

\EFigure{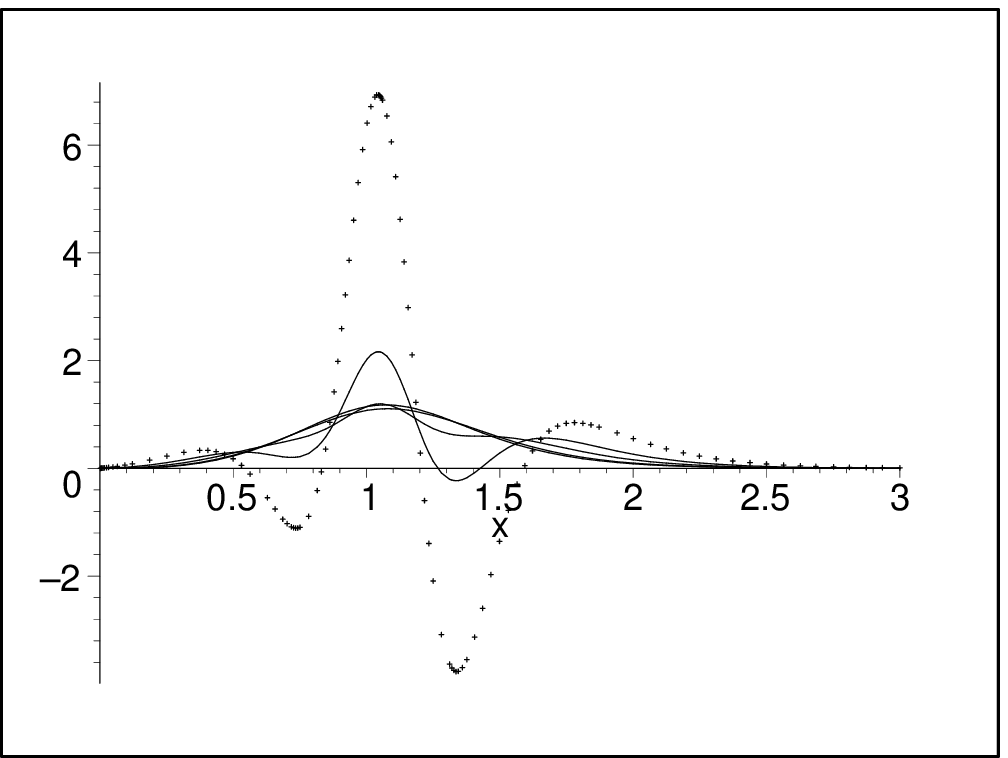}{\label{pic18.eps}Evolution of the energy
distribution of particles,
$d\varepsilon(\tau,x,y)=x^3G(\tau,x,y)$, as a second approximation
at $y=6$. Solid lines from left to right: $\tau=0; 0,01; 0,05;
0,1$, dotted line-$\tau=0,2$.}

\EFigure{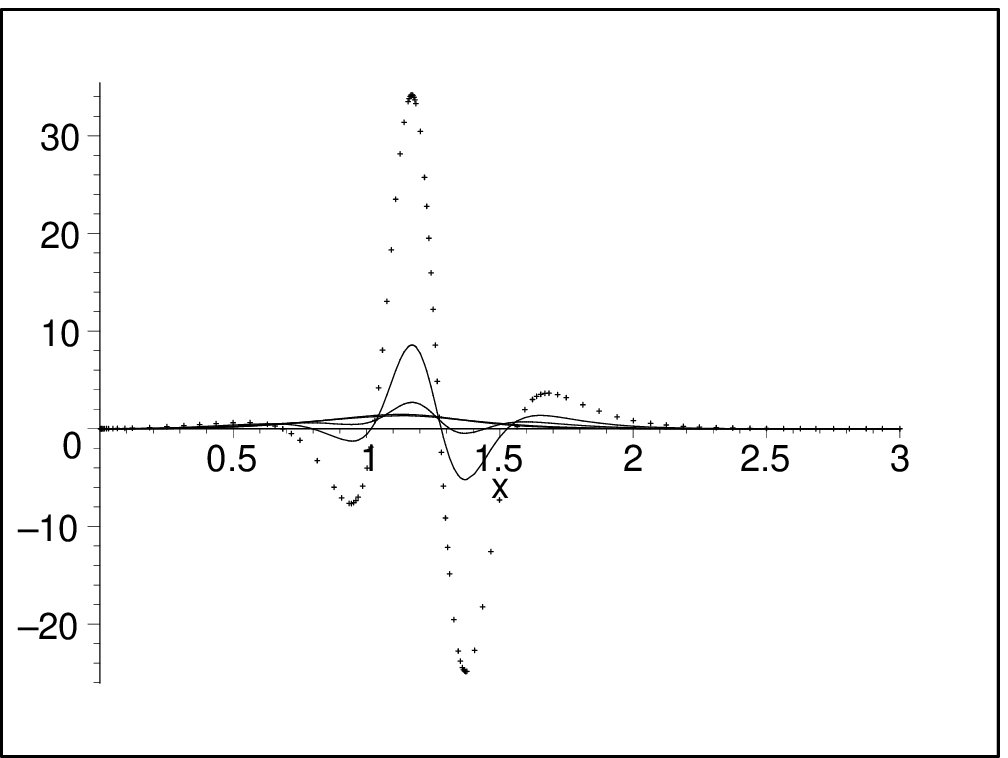}{\label{pic19.eps}Evolution of the energy
distribution of particles,
$d\varepsilon(\tau,x,y)=x^3G(\tau,x,y)$, as a second approximation
at $y=10$. Solid lines from left to right: $\tau=0; 0,01; 0,05;
0,1$, dotted line-$\tau=0,2$.}

\EFigure{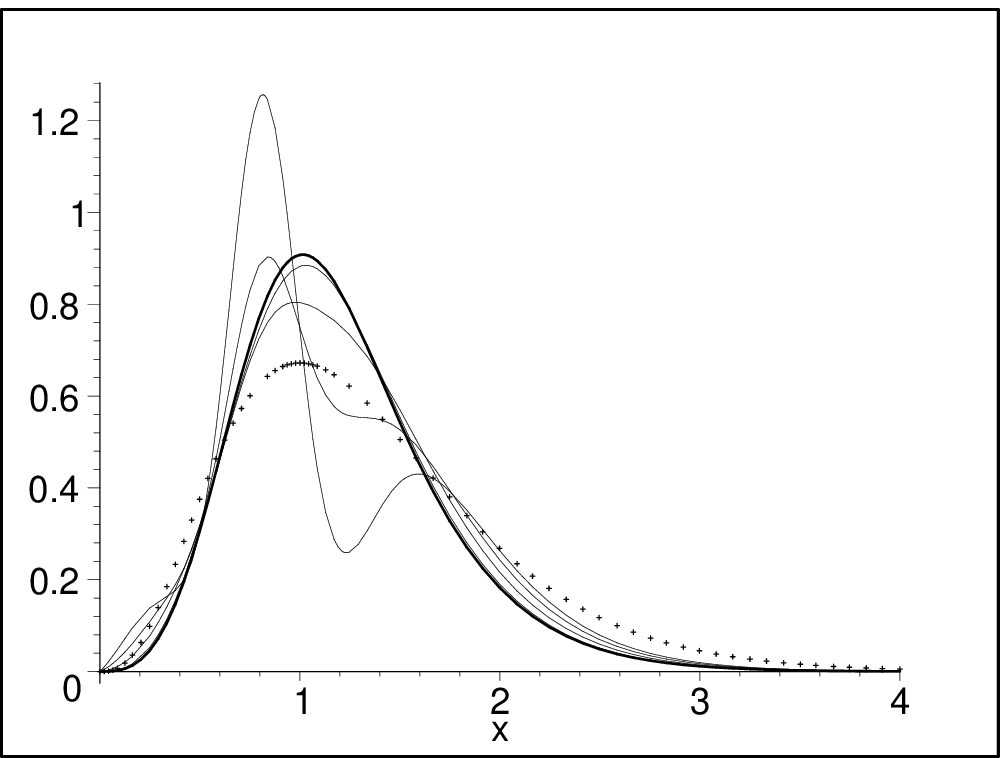}{\label{pic20.eps}Evolution of the energy
distribution of particles,
$d\varepsilon(\tau,x,y)=x^3G(\tau,x,y)$, as a second approximation
at $y=3$. Solid lines from left to right: $\tau=0; 0,01; 0,05;
0,1$, dotted line-$\tau=0,2$.}

\section{Conclusion}

Since particles distribution function is nonnegative by
definition, it is clear that penetration of distribution minimum
to the negative values area is an implication of corrections
smallness conditions failure in the area of concrete energy
values. Examined ap\-p\-ro\-xi\-ma\-ti\-ons sufficiently describe
the global properties of superthermal particles distribution. The
fact of oc\-cur\-ren\-ce of two maximums in superther\-mal
particles distribution is very significant. As a point of view
offered in \cite{Yu1}, \cite{Yu2} superhigh energy particles
origin model, the first of these maximums at a later time can
evolve to equilibrium distribution what certify our calculations,
the second one can give us high-enegry tail of superther\-mal
relic particles.

\vfill

 \pagebreak

\end{document}